\documentclass[aps,prl]{revtex4}
\usepackage{graphicx}

\begin{document}

\def\eps{\varepsilon}
\def\n{\mathrm{n}}
\def\p{\mathrm{p}}

\title{Are pulsar glitches triggered by a superfluid two-stream instability?}

\author{N. Andersson$^*$, G.L. Comer$^\dag$ and R. Prix$^*$}
\affiliation{ $^*$Department of Mathematics, University of
  Southampton, Southampton SO17 1BJ, UK\\ 
$^\dag$Department of Physics, Saint Louis University, St Louis MO63156, USA}

\begin{abstract}
Mature neutron stars are expected to have several superfluid components. Strong evidence for 
this is provided by the glitches that have been observed in dozens of
pulsars. The  underlying idea behind most glitch models
is that, as the neutron star crust spins down due to the emission of
electromagnetic radiation, the superfluid component lags behind
until a critical point is reached and angular momentum is transferred
from the superfluid to the crust, leading to the spin-up associated
with the glitch.  
In this Letter we describe a superfluid analogue of the two-stream 
instability that is well known in plasma physics, and provide evidence
that this instability  is likely to be relevant for neutron
stars. This is a new physical mechanism which  may play a key role in
explaining the glitch mechanism and which could also prove to be
relevant in laboratory experiments on superfluid Helium.  
\end{abstract}

\maketitle

Neutron stars are cosmic laboratories of extreme
physics~\cite{glen}. With a mass of roughly one and a   
half times that of the Sun compressed inside a radius of ten
kilometres, they have density  several times that of nuclear matter,
strong magnetic fields and may spin rapidly.  By now nearly 1300
pulsars (rotating neutron stars) have been observed~\cite{lori}. In
fact,  estimates based on the local population and the supernova rate
suggest that the  total number of neutron stars in the galaxy may be
as large as $2\times10^8$. Despite decades of  theoretical effort the
physics of these compact objects is not well understood.  For example,
there are open issues concerning the pulsar radiation mechanism and
we do not know the detailed supranuclear equation of state. The main
reason for  the many uncertainties is that the description of these
objects requires physics  well beyond our laboratory experience.

Certain aspects of neutron star physics are, however, well established. It is 
clear that the overwhelming majority of these objects must be very cold in the 
sense that their temperatures are significantly below the Fermi temperatures 
of neutrons and protons (typically about $10^{12}$ K). Nuclear physics 
calculations of the transition temperature to superfluidity consistently yield a 
value of the order of $10^9$ K for the neutrons and protons.  A newly born neutron 
star is expected to cool below this temperature within a few weeks to months 
following the supernova explosion.  Thus most neutron stars in our galaxy are 
expected to contain superfluid components~\cite{sauls}. The solid ion crust that forms 
the outer kilometer or so of the star will be permeated by superfluid neutrons, 
while the fluid in the core is expected to contain superfluid neutrons and 
superconducting protons. The deep core may contain exotic phases of matter 
like hyperons and/or deconfined quarks, which may also become
superfluid~\cite{taka,alfo}.  This means that astrophysical neutron
stars provide a manifestation of large scale quantum phenomena,  
and if we want to understand their dynamics we need to allow for the 
presence of, potentially weakly coupled, superfluid components. 

Even though a superfluid is locally irrotational, it can mimic ordinary fluid 
rotation by forming an array of vortices (with typical separation
smaller than $10^{-11}$~m).  In the core of each vortex the
superfluidity is destroyed, the particles are in an ordinary fluid
state and can thus carry non-zero vorticity. Superfluids have zero
viscosity, but dissipative mechanisms can exist when vortices are
present. Particularly important is the direct friction between the
different fluid components (known as "mutual friction"). Another
consequence of the interaction between the fluids is the
non-dissipative entrainment effect. It arises because the bare
neutrons (or protons) are "dressed" by a polarization cloud of
nucleons comprised of both neutrons and protons.  Since both types of
nucleon contribute to the cloud the momentum of the neutrons, say, is
modified so that it is a linear combination of the neutron and proton
currents.  When one of the nucleon fluids begins to flow it will,
through entrainment, induce a momentum in the other. This has the
effect that a portion of the protons (and electrons) will be pulled
along with the superfluid neutrons that surround each vortex. 

Strong observational support for the presence of superfluid components in a 
neutron star is provided by the so-called glitches~\cite{lyne}: sudden spin-up events 
where the observed rotation rate (presumably that of the stars crust) may increase by 
as much as one part in $10^6$. The glitches typically have a relaxation time of the order 
of days to months, which is much longer than can be explained in terms
of normal viscosity. 
 Very soon after the first glitch in the Vela pulsar~\cite{rada,reich}
 was observed in 1969 it was argued~\cite{baym}  
that the long relaxation time indicates the presence of a neutron superfluid. The 
glitch events, and the associated relaxation, are usually explained in
terms of superfluid vortex  
dynamics. This is natural since the superfluid can only change its rotation rate if the 
vortices move. The standard model describes a glitch as an event in which a significant 
number of vortices are suddenly unpinned from the crust
nuclei~\cite{itoh}, angular momentum is  
transferred to the crust and the vortices are eventually repinned. In particular, 
the so-called "vortex creep" model~\cite{cheng} is known to provide an adequate description 
of the glitch relaxation. The actual mechanism that triggers the glitch remains 
unspecified in virtually all models, however. This may be one of the 
most important unresolved issues in this area of research. 

In this Letter we propose that a superfluid two-stream instability (analogous to 
that known to be relevant for electrons streaming through ions in plasma physics~\cite{and}) 
plays a key role in triggering pulsar glitches. The key parameter governing the onset 
of this dynamical instability is the rotational lag between a superfluid component 
and the rest of the star. The existence of such a lag is generally assumed in most 
glitch models and there is observational evidence that glitches occur once this 
lag builds up to a critical level. From a statistical analysis of 48 observed 
glitches in 18 pulsars, Lyne, Shemar and Graham Smith~\cite{lyne} deduce that the glitch 
activity depends linearly on the spin-down rate (as one would expect if a glitch 
is due to the release of a built up rotational difference). They estimate that 
a typical value for the critical velocity is $\Delta
\Omega/\Omega_\p\approx 5\times 10^{-4}$  
(where $\Omega_\p$ and $\Omega_\n$ are the rotation 
rates of the crust and the superfluid neutrons, respectively, and
$\Delta \Omega = \Omega_\n-\Omega_\p$ 
represents the extent to which the superfluid lags behind as the crust spins down).

In a recent paper \cite{acp1} we have studied a superfluid analogue 
of the two-stream instability. As far as we are aware, this mechanism has not previously 
been discussed in this context. Yet, it could turn out to be of extreme importance 
for all superfluid systems that exhibit relative motion. The two-stream instability 
is analogous to the familiar Kelvin-Helmholtz instability~\cite{draz}. Its key distinguishing 
feature is that the two fluids are interpenetrating rather than in contact across an 
interface. In principle, the instability may operate in any system with two interpenetrating 
components moving at different rates. Unstable waves are such that they can be 
associated with a negative energy (in the sense that the energy in the perturbed 
flow is smaller than the energy of the unperturbed system). For this to be possible 
the wave must be such that it moves forwards with respect to one of the fluids, 
but backwards according to an observer riding along with the other fluid. 
Here we consider a neutron star in which the two fluids rotate at constant rates. 
A pulsation mode of the star, 
which is proportional to \mbox{$e^{i(\omega t + m \varphi)}$}, will carry negative energy
if the angular velocity of the mode pattern (the ``pattern speed'') $\sigma_p = -\omega/m$ 
lies in between the two rotation rates, i.e. when
\mbox{$\Omega_\p < \sigma_p<\Omega_\n$}, assuming that the superfluid
neutrons lag behind as the charged component is being spun down
through electromagnetic braking.  

The results described in this Letter were borne out of an effort to understand 
the dynamics of neutron stars with one, or several, superfluid components. The 
simplest models involve two interpenetrating fluids, eg. the superfluid neutrons 
in the inner crust and the core and all charged constituents (the ions in the crust, 
core protons, and all electrons)  and the entrainment effect that acts between 
them~\cite{ac,pca,pr}. We have recently demonstrated \cite{acp1} that the
 two-stream instability may operate in these systems provided that the two fluids 
are coupled, either "chemically" or via the entrainment. In order to provide support 
for the idea that the superfluid two-stream instability is relevant for pulsar 
glitches we consider a model problem of two fluids, allowed to rotate at different 
rates, inside a thin shell. By assuming that the shell is infinitesimally thin we 
may ignore radial motion, which means that the system permits only toroidal 
velocity perturbations. In fact, all oscillation modes of this shell model 
are closely related to the inertial r-modes of rotating single fluid objects~\cite{pap}. 
A somewhat laborious calculation~\cite{acp2} leads to 
the following dispersion relation for oscillation modes of this system
\begin{eqnarray}
&& \left\{ l(l+1) [1-x_\p(1+\eps)](\kappa+\Lambda) -2(1-x_\p) \Lambda
  + x_\p \eps(l-1)(l+2) (1- \Lambda) \right\}  \nonumber \\
&& \times\left\{ l(l+1) [1-\eps](\kappa+1) -2 -  \eps(l-1)(l+2)
(1-\Lambda) \right\} - [l(l+1)]^2 x_\p \eps^2 (\kappa+\Lambda)(\kappa+1) = 0 \ . 
\label{disper}
\end{eqnarray} 
Here we have assumed that the waves are expanded in standard spherical harmonics
$Y_l^m(\theta,\varphi)$ with integer indices $l$ and $|m|\le l$ . Furthermore, 
$x_\p\le0.15$ is the proton fraction, $\kappa=\omega/m\Omega_\p$ 
represents the frequency of oscillation as measured with respect to the 
rotation rate of the charged component, and we have introduced a dimensionless 
measure of the relative rotation $\Lambda=\Omega_\n/\Omega_\p$. The
entrainment effect is represented  
by the coefficient $\eps$, which is expected to lie in the
range $0.4\le\eps\le0.7$  
(see Ref.~\cite{pca} for further discussion). A detailed analysis of
the roots of the above dispersion  
relation shows that complex frequency modes will exist (for any multipole $l$) 
provided that the relative rotation rate $\Lambda$ is sufficiently large. 

As a "physically realistic" example, we consider the case when
$\Delta\Omega/\Omega_\p=5\times10^{-4}$ which corresponds to $\Lambda
= 1.0005$. We then find that we must have $l>65$ in order for the
region of instability to overlap with the anticipated range of the
proton fraction and the entrainment. This indicates that the
instability will operate on length scales shorter than $\pi R/l \approx
500$~m (if we take the shell radius to be $R=10$~km,  
which would be the size of a typical neutron star). The region of instability 
for this rotational lag and $l=100$ is illustrated in Figure~1. Since our 
results only depend on the azimuthal index $m$ through the scaling 
\mbox{$\mathrm{Im}\,\omega= m\Omega_\p\,\mathrm{Im}\,\kappa$},
and the dispersion relation does not depend explicitly on $m$, the fastest growth
 time corresponds to the $m=l$ mode. By analyzing the large $l$ limit of the 
dispersion relation one finds that the fastest possible growth time
for an unstable mode can be approximated by 
\begin{equation}
t \approx { 6.7\times 10^{-2}\over l} \left( { \Delta \Omega \over \Omega_\p} \right)^{-1} 
\left({ P \over 0.1 \mbox{ s}} \right) \mbox{ s}
\label{growth}\end{equation}
This simple formula provides a useful estimate of the fastest growth rate 
of the two-stream instability for different parameter values, and we 
find that for a star rotating at the rate of the Vela pulsar, $P= 89$~ms, 
we would have $t\approx1.2$~s for $l=m=100$. Modes corresponding to higher multipoles 
may grow even faster, but as we will discuss later they are also faster damped
 by viscosity. Interestingly, this predicted growth time is significantly 
shorter than the resolved rise time of a large Vela glitch~\cite{dods}: $t < 40$~s.

\begin{figure}
\centering
\includegraphics[height=5cm,clip]{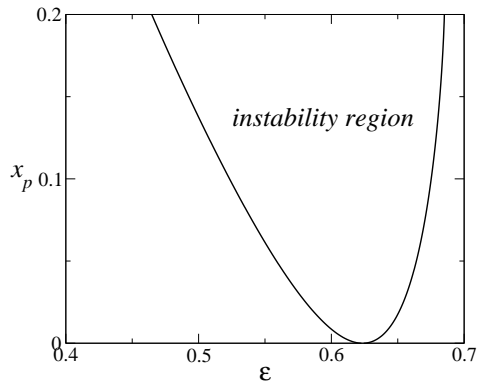}
\caption{The two-stream instability region for the case when the superfluid neutrons 
lag behind the superconducting protons in such a way that $\Delta\Omega/\Omega_\p=5\times10^{-4}$  
($\Lambda= 1.0005$). The 
results are for $l=100$, which means that the unstable modes correspond to oscillations 
on a length scale of a few tens of meters. Similar instability regions exist for all 
$l>65$. Since typical values for a neutron star core are likely to be
such that \mbox{$x_\p < 0.15$} and \mbox{$0.4 \le \eps \le 0.7$}
one would expect the instability to operate in astrophysical neutron stars.}
\end{figure}

Our results suggest that the two-stream instability is generic in dynamical 
superfluids, and could possibly limit the relative flow in any superfluid system. 
Furthermore, the results for the shell model problem demonstrate that entrainment 
provides a sufficiently strong coupling for the instability to operate at 
astrophysically relevant relative flows. We believe that these results indicate 
that the two-stream instability is relevant for neutron stars, and that it may 
be the agent that triggers pulsar glitches. However, much further research is 
required if we are to understand this new mechanism in detail. For example, 
it is crucial to establish that the unstable modes grow faster than the 
relevant dissipation timescales. 
In the case of a superfluid neutron star core the main dissipation mechanisms are 
likely to be mutual friction and shear viscosity due to electron-electron scattering. 
Using estimates due to Mendell (see Eq. 38 in Ref.~\cite{mend}) one can argue that the 
shear viscosity will be the dominant dissipation mechanism for all but the 
lowest values of $l$. To estimate the shear viscosity damping we can use results 
obtained for the secular r-mode instability~\cite{akrev}. Combining
Eq. 43 in Ref.~\cite{ksteg} with the viscosity coefficient for
electron-electron scattering~\cite{akrev} we estimate the viscous
damping time as 
\begin{equation}
t \approx  { 1.2\times10^4 \over l^2 } \left( { 1.4 M_\odot} \over M \right) 
\left({R \over 10 \mbox{ km} }\right)^{5} \left( {T \over 10^7\mbox{
      K} } \right)^2 \mbox{ s} 
\label{tsv}\end{equation}
It is interesting to compare this damping timescale to the growth rate of 
the unstable modes in our shell model.  
We find that, in order to overcome the viscous damping, an unstable mode must correspond to
\begin{equation}
l < 90 \left( {\Delta \Omega / \Omega_\p \over 5\times10^{-4}}\right)
\left({ 0.1 \mbox{ s} \over P} \right) 
\left( { 1.4 M_\odot} \over M \right) 
\left({R \over 10 \mbox{ km} }\right)^{5} \left( {T \over 10^7\mbox{ K} } \right)^2 \ . 
\end{equation}
In the case of the Vela pulsar (the archetypal glitching neutron star) the 
core temperature can be estimated to be $5\times10^7$~K (roughly two orders of magnitude 
higher than the observed surface temperature~\cite{page}). At this temperature  
modes with $l>2500$ or so are likely to be stabilized by shear viscosity.  
Given that our results indicate that the two-stream instability is active for much 
smaller values of $l$, see for example Figure~1, we conclude that dissipation is 
unlikely to suppress the instability in sufficiently young neutron stars. 
Naturally, the situation changes as the star cools further. Our estimates show 
that shear viscosity will suppress all unstable modes if the core temperature 
is below a few times $10^6$~K. Hence, the two-stream instability may not be 
able to overcome viscosity in a sufficiently cold neutron star. This would 
be consistent with the absence of glitches in mature pulsars~\cite{lyne}.  

If the superfluid two-stream instability is, indeed, relevant for pulsar
 glitches then what is its exact role? Much further work is required to 
answer this question, but it is interesting to speculate about the possibilities.
 Most standard models for glitches are based on the idea of catastrophic vortex 
unpinning in the inner crust~\cite{itoh,cheng}. An interesting scenario is the thermally
 induced glitch model due to Link and Epstein~\cite{link}. They show that the deposit of 
$10^{42}$~erg of heat in the crust would be sufficient to induce a large Vela glitch. 
The mechanism that triggers the glitch, eg. by depositing heat in the crust, 
is typically not identified in these models. We propose that the two-stream 
instability may fill this gap in the theory. Of course, glitches need not 
originate in the inner crust. Jones~\cite{pbj} has argued that the vortex pinning is too
 weak to explain the size and frequency of the Vela glitches. Indeed, the observational 
evidence for free precession~\cite{stairs} in PSR B1828-11 indicates the absence of significant
 vortex pinning. If this argument is correct then the large pulsar glitches must be 
due to some mechanism operating in the core fluid. Since our model problem was 
based on equations relevant for the conditions that prevail in a neutron star 
core it is clear that the two-stream instability may equally well serve as a 
trigger mechanism for glitches originating there. One possibility would be 
that the unstable mode grows so large that the superfluid degeneracy is broken, 
which would lead to an immediate coupling to the other normal fluid components. 
The key requirement for the two-stream instability to operate is the presence 
of a rotational lag. Such a lag can build up both when there is a strong coupling 
between the two fluids (i.e. when the vortices are pinned to the crust) and when this coupling 
is weak~\cite{lang}.
 A key issue for future theoretical work on pulsar glitches concerns to what
 extent a rotational lag can build up in various regions of the star. 

Before we conclude this Letter it is worth emphasizing that we expect the 
two-stream instability to be generic in superfluid systems. The two-fluid 
equations that we have employed are analogous to the standard Landau model for 
superfluids, which means that the instability should be relevant also for laboratory systems. 
It is, in fact, not inconceivable that the notion that this 
instability triggers pulsar glitches may be tested using
experiments on superfluid $^4$He. 

\vspace*{0.25cm}
{\em NA and RP acknowledge support from the EU
Programme 'Improving the Human Research Potential and the Socio-Economic
Knowledge Base' (Research Training Network Contract HPRN-CT-2000-00137).
NA acknowledges support from the Leverhulme Trust in the form of a prize 
fellowship. GC acknowledges partial support from NSF grant PHY-0140138.}

\end{document}